\documentstyle[12pt]{article}

\begin{document}
\title{\bf Thermodynamics, Euclidean Gravity,  and Kaluza-Klein Reduction}

\bigskip\bigskip
\bigskip\bigskip

\author{       D. Fursaev $^{\rm a}$
     ~and~
            A. Zelnikov $^{\rm (b,c)}$\thanks{On leave from
        P.N. Lebedev Physics Institute,
        Leninskii prospect 53, Moscow 117 924, Russia. e-mail:
        zelnikov@phys.ualberta.ca}  \\ {}\\
     {$^{\rm a}$\small\it Bogoliubov Laboratory of Theoretical Physics}\\
{\small \it  Joint Institute for Nuclear Research}\\
{\small \it  141980 Dubna, Moscow Region, Russia}\\
{$^{\rm b}$\small\it Theoretical Physics Institute,
University of Alberta,}\\
{\small \it Edmonton, Alberta, Canada T6G 2J1}
\\
{$^{\rm c}$\small\it Yukawa Institute for Theoretical Physics, Kyoto University}\\
{\small \it Kyoto 606-8502, Japan}
}
\maketitle
\begin{abstract}
The aim of this paper is to find out a correspondence between
one-loop effective action $W_E$ defined by means of path integral
in Euclidean gravity and the free energy $F$ obtained by summation
over the modes. The analysis is given for quantum fields on
stationary space-times of a general form. For such problems a
convenient procedure of a "Wick rotation" from Euclidean to
Lorentzian theory becomes quite non-trivial implying transition
from one real section of a complexified space-time manifold to
another. We formulate conditions under which $F$ and $W_E$ can be
connected and establish an explicit relation of these functionals.
Our results are based on the Kaluza-Klein method which enables one
to reduce the problem on a stationary space-time to equivalent
problem on a static space-time in the presence of a gauge
connection. As a by-product, we discover relation between the
asymptotic heat-kernel coefficients of elliptic operators on a $D$
dimensional stationary space-times and the heat-kernel
coefficients of a $D-1$ dimensional elliptic operators with an
Abelian gauge connection.
\end{abstract}

\bigskip

{\it PACS number(s): O4.62.+v, 03.65.Pm, 04.70.Dy}

\bigskip

\baselineskip=.6cm


\newpage

\section{Formulation of the problem}
\setcounter{equation}0

It is a well-known fact that it is easier to work with a quantum
theory on Euclidean space-times than on the Lorentzian ones. Going
to Euclidean signature improves convergence of functional
integrals. This is why Euclidean
methods play an important role in quantum field theory and
especially in quantum gravity including quantum cosmology. Another
application of Euclidean methods is a finite-temperature quantum field theory
where the Euclidean effective action can be interpreted as a
free-energy of  the quantum system with inverse temperature
related to period $\beta$ of
the Euclidean time. In case of black holes this analogue with a finite-temperature theory
appears already at the tree level. As has been first shown by Gibbons and Hawking \cite{GH}, the
classical Euclidean action can be used to infer all thermodynamical properties of a
black hole.  Thus, one may hope that extending the Gibbons-Hawking approach
beyond the tree level approximation is a correct covariant way how to define and
describe canonical ensembles in quantum gravity.

The finite-temperature interpretation of the Euclidean theory is transparent when the
background space-time is static. In this case the free energy
of a canonical ensemble of a Bose field
at the temperature $\beta^{-1}$ is
\begin{equation}\label{i1}
F(\beta)=\beta^{-1} \sum_{\omega} \ln\left(1-e^{-\beta\omega}\right)~~~.
\end{equation}
Here $\omega$ are eigen-values of the time-like Killing operator
$i\partial /\partial_t$. They are the energies of single-particle
states $e^{-i\omega t}\phi_{\omega}(x)$. The corresponding
Euclidean effective action $W_E$ can be defined with the help of
the generalized $\zeta$-function \cite{zeta}
\begin{equation}\label{i2}
W_E(\beta)=-\frac 12 \lim_{\nu \rightarrow 0}{d \over d\nu}
\zeta(\nu|\beta)~~~,
\end{equation}
\begin{equation}\label{i3}
\zeta(\nu|\beta)=\sum_{\omega}\sum_{l=-\infty}^{\infty}
[\sigma_l^2+\omega^2]^{-\nu}~~~,
\end{equation}
where $\sigma_l\equiv (2\pi l)/\beta$.
The $\zeta$-function method provides regularization
of the action.
The relation between (\ref{i1}) and
(\ref{i2}) can be found by
using  the identity (see, e.g., \cite{Fursaev:98})
\begin{equation}\label{f9}
\ln\left(1-e^{-\beta \omega}\right)=-{\beta \omega \over 2}
-\frac 12 \lim_{z\rightarrow 0}{d \over d\nu}\zeta(\nu|\omega,\beta)~~~,
\end{equation}
\begin{equation}\label{f10}
\zeta(\nu|\omega,\beta)=\sum_{l=-\infty}^{\infty}
\left[\sigma_l^2+\omega^2\right]^{-\nu} ~~~.
\end{equation}
It can be proven that (see, e.g., \cite{Allen})
\begin{equation}\label{i4}
W_E(\beta)=\beta (F(\beta)+E_0)
\end{equation}
where $E_0$ corresponds to the energy of vacuum fluctuations.

This simple relation becomes much more involved when
the space-time is stationary. While it does not look difficult to
construct the Euclidean version of a stationary space-time, one
should also find the corresponding transformation of quantum
quantities under the "Wick rotation", what is not trivial in this case.

Any stationary metric can be written in the form
\begin{equation}\label{i5}
ds^2 = -B(dt+a_kdx^k)^2+h_{kj}dx^kdx^j~~~
\end{equation}
where indexes $j,k$ run from 1 to 3 for four-dimensional
space-time, or from $1$ to $D-1$ for $D$-dimensional theory. The
components of the metric (\ref{i5}) depend on spatial coordinates
$x^k$ only. The reference frame corresponding to (\ref{i5}) is
characterized by the acceleration $w_\mu$ and the rotation
$\Omega$ with respect to a local Lorentz frame
\begin{equation}\label{i5x}
w_\mu=\frac 12 (\ln B)_{,\mu}~~~,~~~\Omega^2=\frac 12
A_{\mu\nu}A^{\mu\nu}~~~,
\end{equation}
where in coordinates (\ref{i5}) $A_{ij}=\frac 12 \sqrt{B}
(a_{j,i}-a_{i,j})$ and other components of $A_{\mu\nu}$ are zero.
The local rotation is related to non-diagonal components of the
metric $g_{ti}=-Ba_i$. Presence of these components is an important
distinction from the
static metric which results in mathematical complexity of the
problem.

To see what is the difficulty let us
suppose that an observer whose velocity is parallel to
the Killing vector $\partial/\partial t$ sees a thermal bath of
quanta $e^{-i\omega t}\phi_{\omega}(x)$ at a certain temperature
$T=1/(\beta \sqrt{B})$. Such an equilibrium quantum state is
described by a canonical free energy (\ref{i1}). To make a "Wick
rotation" to Euclidean theory one cannot just replace $t$ in (\ref{i5}) by
an imaginary time but should also change the component of the metric
$g_{ti}$. More precisely, one has to consider (\ref{i5}) as a real
section of a complexified manifold ${\cal M}_c$ and define
Euclidean theory on another real section of ${\cal M}_c$
\footnote{Euclidean section is determined by a concrete physical
problem. It is related to the Killing reference frame where one
defines the state of the thermal equilibrium.}. Thus, the "Wick
rotation" becomes a non-trivial procedure which implies analytical
continuation of the metric. For example, for a rotating black hole
components of metric (\ref{i5}) depend on an additional parameter
$J$, the angular momentum of the black hole, and going to
Euclidean theory requires analytical continuation to imaginary
values of $J$ \cite{GH}. For quantum fields, it means that one has to
consider completely different spectra $\omega$ of single-particle
states in Euclidean and Lorentian theories.

In general, there may be several parameters like $J$ and we will
denote them by a collective symbol $J$. We will assume that the
Euclidean section corresponding to (\ref{i5}) has the metric
\begin{equation}\label{i6}
ds_E^2 = \breve{B}(d\tau+\breve{a}_kdx^k)^2+\breve{h}_{kj}dx^kdx^j~~~
\end{equation}
which is obtained from (\ref{i5}) by replacement $t=-i\tau$ and
analytical continuation of the components to imaginary values of
$J$
\begin{equation}\label{i7}
\breve{h}_{kj}(x;J)=
h_{kj}(x;iJ)~~,~~
\breve{a}_{k}(x;J)=
ia_{k}(x;iJ)~~,~~
\breve{B}(x;J)=
B(x;iJ)~~.
\end{equation}
Let us denote such Euclidean manifold with metric (\ref{i6}) by
${\cal M}_E$ and assume that $\tau$ is a periodic coordinate with
period $\beta$. The Euclidean effective action $W_E(\beta,J)$ can
be defined by a path integral of the corresponding theory on
${\cal M}_E$. For free fields or in one-loop approximation it reduces to determinants of
Euclidean wave operators. In what follows we consider a regularized
action defined as
\begin{equation}\label{i8} W_E(\beta,J)
=-\frac 12 \lim_{\nu\rightarrow 0}{d \over d\nu} \zeta(\nu|\beta,J)~~~,
\end{equation}
\begin{equation}\label{i9}
\zeta(\nu|\beta,J)=\sum_{\Lambda}
\left[\varrho^2\Lambda\right]^{-\nu}~~~,
\end{equation}
where $\Lambda$ are eigen-values of the corresponding wave
operators on ${\cal M}_E$; $\varrho$ is a dimensional parameter
related to regularization. Functional (\ref{i8}) does not have an
immediate physical meaning. To make the relation to the physical
space-time one has to analytically continue (\ref{i8}) back to
some real values of $J$ (these values are to be in an interval
corresponding to the rotation with the velocity smaller than the
velocity of light). This results in a new functional
\begin{equation}\label{i10}
\tilde{W}_E(\beta,J)
=W_E(\beta,-iJ)~~~.
\end{equation}
This functional divided by $\beta$ is also frequently interpreted as a
free energy of a quantum field.
Thus, the Euclidean method includes the three steps:
definition of the Euclidean section ${\cal M}_E$, computation of the Euclidean action
$W_E(\beta,J)$ on ${\cal M}_E$, and inverse "Wick rotation" from $W_E(\beta,J)$  to
$\tilde{W}_E(\beta,J)$.

\bigskip

The problem which is studied in the present paper is the relation
between the one-loop effective action $\tilde{W}_E(\beta,J)$ and the
canonical free energy $F(\beta,J)$ defined by (\ref{i1}) on a
stationary space-time. The motivation for this study is simple.
Relation (\ref{f9}) cannot be directly used for stationary metrics
and we want to understand how, in this case, the Euclidean method
reproduces the summation over the modes $\omega$.

\bigskip

Our analysis will be based on a new approach to quantum effects in
case of non-vanishing rotation. Its idea, which was suggested in
\cite{FF:2000} and elaborated in \cite{Fursaev:2000}, is to reduce
computations on a stationary space-time to a fiducial problem on a
static space-time but in the presence of a fiducial (Abelian) gauge
connection $a_kdx^k$. The transition from one problem to another
and appearance of the gauge field are analogous to the reduction
in the Kaluza-Klein theory. That is why the method is called
Kaluza-Klein method (KK-method for simplicity). This method is
most suitable for our purpose because we want to make computations
similar to the static case where the relation between Euclidean
and canonical approaches is well established. This also has
another use. We rederive with the help of the Euclidean approach
some results found in \cite{Fursaev:2000} and, hence, give
an independent check of the KK-method itself. We also present a
number of new results. One of them is a zeta-function definition
of the vacuum energy of a field on a stationary space-time.
The other result is a
Kaluza-Klein reduction formula for the coefficients of the
asymptotic heat kernel expansion. The formula relates  heat-kernel
coefficients of elliptic second order operators on a $D$
dimensional stationary space-time to heat kernel coefficients of
elliptic operators on a $D-1$ dimensional space-time with a gauge connection.

\bigskip

The plan of the paper is as follows. In the next Section  we describe
the KK method in Lorentzian and Euclidean theories. In Section 3
we formulate conditions which enable one to make a "Wick
rotation" and relate  $\tilde{W}_E$ and $F$ on an arbitrary
stationary space-time. We prove that under these conditions the
relation between both functionals has the same form as
(\ref{i4}). To support our analysis in Section 4 we re-derive
by methods of the Euclidean theory high temperature asymptotics
previously found in \cite{Fursaev:2000}. Section 5 is devoted to
new results. Here we discuss  the vacuum energy and Kaluza-Klein
reduction formulas for heat-kernel coefficients. Section 7
contains summary and discussion of the results. In Appendix we demonstrate
explicitly how the reduction formulas work for first heat-kernel
coefficients.

\section{The KK method}
\setcounter{equation}0

Let us first remind results of \cite{Fursaev:2000}. Consider a
wave equation for  a free  scalar field
\begin{equation}\label{kk1}
(-\nabla^\mu\nabla_\mu+V)\phi=0~~.
\end{equation}
We choose a Killing reference frame where metric is given by
(\ref{i5}). We are interested in solutions  of (\ref{kk1}) which
describe single-particle states with a certain energy $\omega$
with respect to time $t$. These solutions have the following form
$e^{-i\omega t}\phi_\omega(x)$ where $\phi_\omega(x)$ are
solutions of a $D-1$ dimensional differential equation.

In the Killing frame (\ref{i5})
the wave operator can be represented as
\begin{equation}\label{kk2}
\nabla^\mu\nabla_\mu=-{1 \over B}
(\xi^\mu\nabla_\mu)^2+{1 \over 2B} B^{,\nu}\nabla_\nu +
h^{\mu \nu}\nabla_\mu \nabla_\nu=
-{1 \over B}\partial_t^2-{1 \over 2B} B_{,i}h^{ij}D_j+
h^{ij}D_iD_j~~~,
\end{equation}
where $D_j=\partial_j-a_j\partial_t$.
It is easy to see now that the wave operator (\ref{kk2}) can be written
in another form as
\begin{equation}\label{kk3}
\nabla^\mu\nabla_\mu=\tilde{g}^{\mu\nu}D_\mu D_\nu~~~,
\end{equation}
\begin{equation}\label{kk4}
D_\mu=\tilde{\nabla}_\mu-a_\mu \partial_t~~~,
\end{equation}
where $a_\mu dx^\mu=a_i dx^i$. The covariant derivatives
$\tilde{\nabla}_\mu$
are defined on some static space-time $\tilde{\cal M}$
with the metric
\begin{equation}\label{kk5}
d\tilde{s}^2=\tilde{g}_{\mu\nu}dx^\mu dx^\nu=-Bdt^2+
h_{jk}dx^j dx^k~~~.
\end{equation}
This metric is obtained from (\ref{i5})
by imposing condition $a_i=0$.
$\tilde{\cal M}$ and $a_\mu$ are called the fiducial
space-time and the fiducial gauge potential, respectively.

Following to \cite{Fursaev:2000} we can introduce a scalar field
$\phi^{(\lambda)}$ on
$\tilde{\cal M}$
as a solution to equation
\begin{equation}\label{kk6}
(-\tilde{g}^{\mu\nu} (\tilde{\nabla}_\mu+i\lambda a_\mu)
(\tilde{\nabla}_\nu+i\lambda a_\nu)+m^2)\phi^{(\lambda)}=0~~~
\end{equation}
and consider a single-particle excitation
$\phi^{(\lambda)}_\omega(t,x^i)=e^{-i\omega
t}\phi^{(\lambda)}_\omega(x^i)$ of this field. By comparing
(\ref{kk6}) with (\ref{kk1})--(\ref{kk3}) one can identify
$\phi^{(\omega)}_\omega$ with a physical single-particle
excitation with the energy $\omega$, i.e. with a solution to
(\ref{kk1}). This enables one to go from a problem on the
stationary space-time $\cal M$ to a problem on the fiducial {\it
static} background $\tilde{\cal M}$ and in an external gauge
potential $a_\mu$. Equation (\ref{kk6}) reduces to the eigen-value
problem
\begin{equation}\label{kk7}
H^2(\lambda)\phi^{(\lambda)}_{\omega}(x^i)=\omega^2
\phi_{\omega}^{(\lambda)}(x^i)~~~,
\end{equation}
where $H(\lambda)$ has the meaning of a relativistic single-particle
Hamiltonian for the field $\phi^{(\lambda)}_\omega(x^i)$.

If we consider a system with a continuous spectrum definition
(\ref{i1}) of the free energy becomes\footnote{In
\cite{Fursaev:2000} the denity of levels $\Phi(\omega)$  is
denoted by $dn(\omega)/d\omega$. In the present work the different
notation is chosen  for the convenience.}
\begin{equation}\label{kk8}
F(\beta)=\beta^{-1}\int_{\mu}^{\infty} \Phi(\omega) d\omega
\ln\left(1-e^{-\beta\omega}\right)~~~.
\end{equation}
Here $\mu$ is the mass gap and
$\Phi(\omega)d\omega$ is the number of
normalized solutions $e^{-i\omega t}\phi_\omega(x)$ of
equation (\ref{kk1}) in the interval of energies
$(\omega,\omega+d\omega)$. We assume that $\Phi(\omega)$ is
a regularized quantity. This regularization can be introduced
by cutting off the integrations over the space, which is equivalent
to restricting the size of the system.

Suppose that
$\Phi(\omega;\lambda)$
is the properly regularized
density of energy levels $\omega$ of the
corresponding fiducial problem
(\ref{kk6}).  From mathematical point of view
$\Phi(\omega;\lambda)$ is the spectral measure of
the operator (\ref{kk7}) so that one can write
\begin{equation}\label{kk9}
\int_{\mu}^{\infty} \Phi(\omega;\lambda) d\omega e^{-t\omega^2}
=\mbox{Tr}~e^{-tH^2(\lambda)}~~~.
\end{equation}
The trace of $H^2(\lambda)$ has to be understood in the above
sense, as a regularized quantity. The physical density of levels has been found in \cite{Fursaev:2000}.
It is given by the relation
\begin{equation}\label{kk11}
\Phi(\omega)= \left.\Psi(\omega;\lambda)\right|_{\lambda=\omega}~~~,
\end{equation}
where $\Psi(\omega;\lambda)$ is defined as
\begin{equation}\label{kk10}
\int_{\mu}^{\infty} \Psi(\omega;\lambda) d\omega e^{-t\omega^2}
=\left(1+{1 \over 2\lambda t}\partial_\lambda\right)
\mbox{Tr}~e^{-tH^2(\lambda)}~~~.
\end{equation}
One can prove that the latter relation is equivalent to
\begin{equation}\label{kk12}
\Psi(\omega;\lambda)=\Phi(\omega;\lambda) +{\omega \over
\lambda}\int_{\mu}^{\omega}
\partial_\lambda \Phi(\sigma;\lambda)d\sigma~~~.
\end{equation}
It should be noted that $\Phi(\omega)$ does not coincide with $\Phi(\omega;\omega)$, as
one could naively expect. The distinction of these two quantities is
in different forms of the inner products in the physical and fiducial problems, see \cite{Fursaev:2000}
for the details.

The idea behind the KK method is to compute the fiducial
density $\Phi(\omega;\lambda)$ by using the heat kernel of
the operator $H^2(\lambda)$ and then get the physical
density $\Phi(\omega)$ with the help of (\ref{kk11})--(\ref{kk12}).

\bigskip

Let us consider the Euclidean formulation of the
corresponding problem. We are interested now in functional (\ref{i8})
on Euclidean manifold (\ref{i6}) where $\tau$ is a periodic coordinate
with period $\beta$. The $\zeta$-function (\ref{i9})
is determined by the eigen-values of the Euclidean wave operator
on ${\cal M}_E$
\begin{equation}\label{kk13}
(-\nabla^\mu\nabla_\mu+V)\phi_\Lambda=\Lambda \phi_\Lambda~~.
\end{equation}
Because of the isometry of the background manifold $\phi_\Lambda$
in (\ref{kk13}) are eigen-vectors of the operator
$i\partial_\tau$
\begin{equation}\label{kk14}
\phi_{\Lambda}(\tau,x)=e^{i\sigma_l\tau}
\phi_{\Lambda}(x)~~,
\end{equation}
where $\sigma_l=(2\pi l)/\beta$,  $l=0,\pm 1, \pm 2,...$.
For these modes (\ref{kk13}) takes the form
\begin{equation}\label{kk15}
(-\breve{g}^{\mu\nu}_E (\breve{\nabla}_\mu+i\sigma_l \breve{a}_\mu)
(\breve{\nabla}_\nu+i\sigma_l
\breve{a}_\nu)+V)
\phi_{\Lambda}=\Lambda \phi_{\Lambda}~~.
\end{equation}
Here the metric and the connections correspond to a fiducial
static Euclidean space-time $\tilde{\cal M}_E$
\begin{equation}\label{kk16}
d\breve{s}^2_E=(\breve{g}_E)_{\mu\nu}dx^\mu dx^\nu=\breve{B}d\tau^2+
\breve{h}_{jk}dx^j dx^k~~~.
\end{equation}
Thus, the Euclidean problem on a stationary background also can be
reformulated as a theory on a static background in the presence
of a fiducial gauge connection.  The distinction from the
Lorentzian theory is that the fiducial charges $\sigma_l$ are quantized
because the Euclidean time coordinate $\tau$ is compact.

Without loss of generality we can
suppose that $\breve{B}=1$ in (\ref{kk16}). One can always
use conformal transformation to bring the metric to this
form. The result of this transformation is an anomalous addition to the
Euclidean action. This addition is proportional
to $\beta$ \cite{Dowker:89} and changes only the
vacuum energy $\breve{E}_0$ (see Section 5 below).
If $\breve{B}=1$ the eigen-value problem can be written as
\begin{equation}\label{kk17}
(\sigma_l^2+\breve{H}^2(\sigma_l))\phi_{\Lambda}
=\Lambda\phi_\Lambda~~,
\end{equation}
\begin{equation}\label{kk18}
\breve{H}^2(\sigma_l)=-\breve{h}^{jk}(\breve{\nabla}_j+
i\sigma_l \breve{a}_j)
(\breve{\nabla}_k+i\sigma_l
\breve{a}_k)+V~~.
\end{equation}
Operator $\breve{H}^2(\sigma_l)$ is analogous to the Lorentzian
operator $H^2(\lambda)$. One may say that
$\breve{H}^2(\sigma_l)$ is obtained from $H^2(\lambda)$ under
analytical continuation (\ref{i7}) of the metric and replacement $\lambda$ by
$i\sigma_l$.
Given the solution of the eigen-value problem
\begin{equation}\label{kk19}
\breve{H}^2(\sigma_l)\phi^{(\sigma_l)}_{\omega}(x^i)=\omega^2
\phi^{(\sigma_l)}_{\omega}(x^i)~~~
\end{equation}
the solution of (\ref{kk13}) can be represented as
\begin{equation}\label{kk20}
\phi_{\Lambda}(\tau,x)=e^{i\sigma_l\tau}\phi^{(\sigma_l)}_{\omega}(x^i)~~,~~
\Lambda=\sigma_l^2+\omega^2~~~.
\end{equation}
Let us denote $\breve{\Phi}(\omega;\sigma_l)$ the (regularized) spectral density
of $\breve{H}^2(\sigma_l)$. Then the Euclidean
effective action can be written as (see (\ref{i8}), (\ref{i9}))
\begin{equation}\label{kk21}
W_E(\beta)
=-\frac 12 \lim_{\nu\rightarrow 0}{d \over d\nu}
\zeta(\nu|\beta)~~~,
\end{equation}
\begin{equation}\label{kk22}
\zeta(\nu|\beta)=\varrho^{-2\nu}
\int_{\mu}^{\infty} d\omega \sum_{\sigma_l}
\breve{\Phi}(\omega;\sigma_l)
(\sigma_l^2+\omega^2)^{-\nu}~~~.
\end{equation}
Here we assume that $\breve{H}^2(\sigma_l)$ have a positive mass gap $\mu$ which does not depend on
$\sigma_l$.
Although (\ref{i9}), (\ref{kk22}) make sense when $Re~\nu>D/2-1$
one can consider analytical continuation to show that $\zeta(\nu|\beta)$
is a meromorphic function of $\nu$
with simple poles on the real axis (see, e.g., \cite{BCVZ:96}).
In particular, this function is regular near $\nu=0$ so that one can use (\ref{kk21}).
Definitions (\ref{kk8}), (\ref{kk11}), (\ref{kk12})  and
(\ref{kk21}), (\ref{kk22}) will be our starting point.

\section{Euclidean action and free energy}
\setcounter{equation}0

We start with definition of the Euclidean action, Eqs. (\ref{kk21})
and (\ref{kk22}). The $\zeta$-function can be rewritten
by using the Cauchy theorem as
\begin{equation}\label{ec1}
\zeta(\nu|\beta)={\varrho^{-2\nu} \over 2\pi i}
\int_{\mu}^{\infty} d\omega \int_C dz \sum_{\sigma_l}
{\breve{\Phi}(\omega;z) \over z-\sigma_l}
(z^2+\omega^2)^{-\nu}~~~,
\end{equation}
where the contour $C$ consists of two parallel lines, $C_+$ and $C_-$,
in the complex
plane. $C_+$ goes from $(i\epsilon+\infty)$
to $(i\epsilon-\infty)$ and $C_-$  goes from
$(-i\epsilon-\infty)$ to
$(-i\epsilon+\infty)$. Here $\epsilon$ is a small positive
parameter such that $\epsilon<\mu$.   We consider (\ref{ec1}) at $Re~\nu> D/2-1$
and assume that $D\geq 2$.
The sum over $\sigma_l$ in (\ref{ec1}) can be performed
\begin{equation}\label{ec2}
\zeta(\nu|\beta)={\varrho^{-2\nu}\beta \over 4\pi i}
\int_{\mu}^{\infty} d\omega \int_C dz \cot \left({\beta z \over 2}
\right)
\breve{\Phi}(\omega;z)
(z^2+\omega^2)^{-\nu}~~~.
\end{equation}
Note that the spectral density $\breve{\Phi}(\omega;z)$
is an even function of the "charge" $z$. Hence
integrations over $C_+$ and $C_-$ in (\ref{ec2}) coincide
and one can write (\ref{ec2}) as a twice of the integral
over $C_+$.
To proceed let us use the identity
$$
\cot\left({\beta z \over 2}\right)=
{2 \over \beta}{d \over dz}
\ln\left(1-e^{i\beta z}\right)-i$$
which enables one to write
\begin{equation}\label{ec3a}
\zeta(\nu|\beta)=\beta\zeta_0(\nu)+
\zeta_T(\nu|\beta)~~~,
\end{equation}
\begin{equation}\label{ec3b}
\zeta_0(\nu)={\varrho^{-2\nu} \over \pi }
\int_{\mu}^{\infty} d\omega \int_{0}^{\infty}
\breve{\Phi}(\omega;x) (x^2+\omega^2)^{-\nu}dx~~~,
\end{equation}
\begin{equation}\label{ec21a}
\zeta_T(\nu|\beta)={\varrho^{-2\nu} \over  \pi i}
\int_{\mu}^{\infty} d\omega \int_{C_+} dz
{d \over dz}
\ln\left(1-e^{i\beta z}\right)
\breve{\Phi}(\omega;z)
(z^2+\omega^2)^{-\nu}~~~.
\end{equation}
Function  (\ref{ec21a}) represents purely thermal part which vanishes
at the zero temperatue because of small positive imaginary part of $z$
in $e^{i\beta z}$. It means that the only quantity responsible for
the vacuum energy is
$\zeta_0(\nu)$.
Let us now deform $C_+$ in (\ref{ec21a}) so that to make
the integrand exponentially small at large $z$ due to the factor $e^{i\beta z}$
\footnote{
The following arguments show that the spectral density $\breve{\Phi}(\omega;z)$ itself cannot increase
at large $z$ and so  the presence of factor $e^{i\beta z}$ is  enough
to ensure convergence of (\ref{ec21a}). Indeed,
the parameter $z$ is a "charge"
of a particle described by the "Hamiltonian" $\breve{H}(z)$.
The generalized momentum of such
particle is
$p_i=i(\breve{\nabla}_i+iz\breve{a}_i)$ and $|p_i|$ is large at large $z$.
From the point of view of quantum mechanics,
$\breve{\Phi}(\omega;z)d\omega$ is the probability
for a particle to have an
energy in the interval $(\omega,\omega+d\omega)$.
This probability should be small when $\omega$ is finite
while the momentum is arbitrary large.}.
After that we can integrate by parts to get
\begin{equation}\label{ec3}
\zeta_T(\nu|\beta)=
{\varrho^{-2\nu} \over \pi i}
\int_{\mu}^{\infty} d\omega \int_{C_+} dz
\ln\left(1-e^{i\beta z}\right)
\left[{2\nu z \breve{\Phi}(\omega;z) \over
(z^2+\omega^2)^{\nu+1}}-
{\partial_z\breve{\Phi}(\omega;z) \over
(z^2+\omega^2)^{\nu}}
\right]
~~.
\end{equation}
We can then integrate
the first term in the square brackets in the r.h.s. of (\ref{ec3})
by parts over $\omega$
and get
\begin{equation}\label{ec4}
\zeta_T(\nu|\beta)=
{\nu \varrho^{-2\nu}\over \pi i}
\int_{\mu}^{\infty} d\omega \int_{C_+} dz
{2z \over (z^2+\omega^2)^{\nu+1}}
\ln\left(1-e^{i\beta z}\right)
\breve{\Psi}(\omega;z) ~~~,
\end{equation}
\begin{equation}\label{ec5}
\breve{\Psi}(\omega;z)=\breve{\Phi}(\omega;z)-{\omega \over z} \int_\mu^\omega {d \over dz}
\breve{\Phi}(\sigma;z)d\sigma~~~.
\end{equation}
The boundary terms vanish at $\mbox{Re}~\nu>D/2-1$ because
$\partial_z\breve{\Phi}(\omega;z)\sim \omega^{D-4}$ at large $\omega$
(this follows from (\ref{ec5}) and
(\ref{as4}), see below).
We can now represent the Euclidean action in the following
form (see (\ref{kk21}), (\ref{kk22}))
\begin{equation}\label{ec6}
W_E(\beta)
=\beta (\breve{F}(\beta) - \breve{E}_0)~~~,
\end{equation}
\begin{equation}\label{ec7}
\breve{F}(\beta)=-\frac 12\lim_{\nu\rightarrow 0}{d \over d\nu}
\zeta_T(\nu|\beta)~~~,
\end{equation}
\begin{equation}\label{ec9}
\breve{E}_0=-\frac 12\lim_{\nu\rightarrow 0}{d \over d\nu}\zeta_0(\nu)
\end{equation}
where $\zeta_T(\nu|\beta)$ and $\zeta_0(\nu)$ are given by (\ref{ec4}), (\ref{ec3b}), respectively.
The quantity $\breve{E}_0$ which corresponds to the vacuum energy will be
discussed in Section 5.
We focus  on $\breve{F}(\beta)$.
To compute it let us note that $\zeta_T(\nu|\beta)$ has a form $\zeta(\nu|\beta)=\nu f(\nu|\beta)$,
see (\ref{ec4}). To find $f(\nu|\beta)$ at $\nu=0$ we make our first assumption.

\bigskip
\noindent
{\it Assumption 1.} For any $\omega$ density $\breve{\Psi}(\omega,z)$ can be analytically continued to complex $z$
and is an entire function of $z$ in the upper half of the complex plane.

\bigskip
\noindent
Let us add to the
contour $C_+$ a large semicircle lying in the upper half of the complex plane and make a
closed contour. Because of the exponent $e^{i\beta z}$ in the logarithm in
(\ref{ec4}) the integration over the semicircle vanishes when its
radius goes to infinity. Our assumption guarantees that  function
$f(\nu|\beta)$ is finite at $\nu=0$ and by the Cauchy theorem its value
is determined only by the residue at
$z=i\omega$.  The result for (\ref{ec7}) looks as follows
\begin{equation}\label{ec11}
\breve{F}(\beta)=-\frac 12f(0|\beta)={1 \over \beta}
\int_{\mu}^{\infty}
\breve{\Phi}(\omega) d\omega
\ln\left(1-e^{-\beta \omega}\right)~~~,
\end{equation}
\begin{equation}\label{ec12}
\breve{\Phi}(\omega)
=\left.\breve{\Psi}(\omega;z)\right|_{z=i\omega}~~~.
\end{equation}
Functional (\ref{ec11}) has a form
of the free energy of a system with the density
of levels (\ref{ec12}).
Similarity of $\breve{\Phi}(\omega)$ and the physical density of levels $\Phi(\omega)$
defined by (\ref{kk11}), (\ref{kk12}) is obvious.  We now restrict ourselves by
problems where transition from Euclidean ${\cal M}_E$ to Lorentzian space-time $\cal M$
is determined by analytical continuation
of a set of parameters $J$ of the metric, as was discussed in Introduction.
We also make a second assumption

\bigskip
\noindent
{\it Assumption 2.} The density $\breve{\Psi}(\omega;\lambda)$, as a function of parameters
$J$, can be analytically continued to complex $J$ in such a way that the following equality holds
true
\begin{equation}\label{ec13}
\breve{\Psi}(\omega;i\lambda|-iJ)=\Psi(\omega;\lambda|J)~~~
\end{equation}
for $J$ and $\lambda$ real.

\bigskip
\noindent
In fact, condition (\ref{ec13}) determines a "Wick rotation"
in a quantum theory on a stationary
gravitational background.
It implies that spectra of
$\breve{H}^2(\lambda)$ and $H^2(\lambda)$ are related by
the analytical continuation.

Given equations (\ref{ec11})--(\ref{ec13})
one concludes that
$\breve{\Phi}(\omega)$ and
$\breve{F}(\beta)$
coincide
after the "Wick rotation"
with the physical density of levels and the free energy, respectively,
\begin{equation}\label{ec14a}
\breve{\Phi}(\omega|-iJ)=\Phi(\omega|J)~~~,
\end{equation}
\begin{equation}\label{ec14}
\breve{F}(\beta|-iJ)=F(\beta|J)~~~.
\end{equation}
This is the main result of this Section which extends the relation between statistical mechanics
and Euclidean theory to the most general case of stationary gravitational fields.
It should be noted, however, that it is not a full proof because (\ref{ec14}) is based on
the two assumptions about analytical properties of the spectral density
of elliptic operators. Thus, one should check whether these assumptions are really fulfilled.
We postpone this issue till the next Section.

We conclude this Section with comments regarding the case
when operators $\breve{H}^2(\sigma_l)$ have  discrete spectra.
The $\zeta$-function (\ref{kk22}) of the Euclidean wave operator
can be  written as
\begin{equation}\label{ec23}
\zeta(\nu|\beta)=\varrho^{-2\nu}\sum_{\sigma_l}\sum_\omega
(\sigma_l^2+\breve{\omega}^2(\sigma_l))^{-\nu}~~~,
\end{equation}
where the sum is taken over
all  eigen-values $\breve{\omega}^2(\sigma_l)$ of $\breve{H}^2(\sigma_l)$
(some eigen-values may coincide). By analogy with the
Lorentzian theory let us consider a one-parameter
family of operators $\breve{H}^2(\lambda)$
such that at $\lambda=\sigma_l$
one gets operators used in definition (\ref{ec23}). Denote their eigen-values by
$\breve{\omega}^2(\lambda)$.
Let us make the following assumptions:

1) function $f(z)=(\breve{\omega}^2(z)+z^2)^{-1}$ can be analytically continued in $z$ to
the upper complex plane where it is a meromorphic function with simple
poles which lie strictly on the axis $Re~z=0$;

2) the "Wick rotation" to Lorentzian theory is determined by analytical continuation
of a set of parametrs $J$ in such a way that under this continuation eigen-values
of $\breve{H}^2(\lambda)$ transform into eigen-values of the Lorentzian
operator $H^2(\lambda)$, i.e.
\begin{equation}\label{corr}
\breve{\omega}^2(i\lambda|-iJ)=\omega^2(\lambda|J)~~~.
\end{equation}

\noindent
These conditions are analogous to the two assumptions made in the case
of continuous spectrum.
We now replace the sum over $\sigma_l$ in (\ref{ec23}) by the integration in the complex plane
and decompose $\zeta(\nu|\beta)$ into
two parts, as in (\ref{ec3a}),
\begin{equation}\label{ec24}
\zeta_0(\nu)={\varrho^{-2\nu} \over \pi }
\int_0^\infty dx \sum_{\omega}d_\omega
(x^2+\breve{\omega}^2(x))^{-\nu}~~~.
\end{equation}
\begin{equation}\label{ec25}
\zeta_T(\nu|\beta)=
{\varrho^{-2\nu} \over \pi i}
\int_{C_+} dz  \sum_\omega d_\omega
\ln\left(1-e^{i\beta z}\right)
{2\nu z  \over (z^2+\breve{\omega}^2(z))^{\nu+1}}  \left[1+
{1 \over 2z}
{\partial \breve{\omega}^2(z)\over \partial z}\right]
~~.
\end{equation}
(To get (\ref{ec25}) we have integrated by parts.)
Analogously, as in (\ref{ec6}) the Euclidean action can be divided onto the vacuum part $\breve{E}_0$
and the thermal part $\breve{F}(\beta)$.
The vacuum energy $\breve{E}_0$ is determined by (\ref{ec9})
with $\zeta_0(\nu)$ given by (\ref{ec24}).
The free energy $\breve{F}(\beta)$ is determined by (\ref{ec7}).
Function $\zeta_T(\nu|\beta)$ near $\nu=0$
can be evaluated by using Cauchy theorem and the condition about the poles.
By taking into account that near a pole  $z=iz_i$
$$
{1 \over z^2+\breve{\omega}^2(z)}
\simeq {1 \over (2iz_i+\partial_z(\breve{\omega}^2(iz_i))(z-iz_i)}
$$
and in the upper complex plane $z_i=\breve{\omega}(iz_i)>0$ we get
\begin{equation}\label{ec26}
\breve{F}(\beta)={1 \over \beta}
\sum_{\breve{\omega}(iz_i)}
\ln\left(1-e^{-\beta \breve{\omega}(iz_i)}\right)~~~.
\end{equation}
Finally, the second assumption guarantees that after the "Wick rotation"
(\ref{ec26}) coincides with the physical free energy
\begin{equation}\label{ec27}
F(\beta)={1 \over \beta}
\sum_\omega
\ln\left(1-e^{-\beta \omega}\right)~~~,
\end{equation}
where the sum is taken over all solutions of equation
$\omega^2(z)=z^2$.

\section{Asymptotics}
\setcounter{equation}0

The relation between the free energy and the effective action
proved in the previous Section is based on assumption
about analytical properties of the spectral density.
It is difficult to give a rigorous proof of (\ref{ec13})
because the explicit form of $\Phi(\omega;\lambda)$ is not
known in general. Here we demonstrate that our assumption
holds true at least asymptotically in the limit of
high frequencies. In this limit the spectral density
is a local functional of the background geometry whose
form is determined by the asymptotic expansion of
the heat-kernel of the operator $H^2(\lambda)$
\begin{equation}\label{as1}
\mbox{Tr} e^{-tH^2(\lambda)}
\simeq {1 \over (4\pi t)^{(D-1)/2}}
\sum_{n=0}^\infty\left[a_n(\lambda) t^n
+ b_n(\lambda) t^{n+\frac 12}\right]~~~
\end{equation}
at small
values of $t$.
Here $a_n$ and $b_n$
are the standard heat kernel coefficients, $n=0,1,2,...$.
On manifolds without boundaries $b_n=0$.
Coefficients $a_n(\lambda)$ and
$b_n(\lambda)$ are even functions of $\lambda$ because
the theory is $U(1)$ invariant and the heat coefficients
are even functions of charges.
The density of levels at high frequencies  can be found
from (\ref{as1}) by using the inverse Laplace transform.
By neglecting the mass gap
and treating $D$ as a complex parameter to avoid infrared
singularities one gets \cite{Fursaev:2000}
\footnote{For discussion of analogous expansion
for integer $D$ see \cite{CVZ:89}.}
\begin{equation}\label{as2}
\Phi(\omega;\lambda)\simeq
{2 \omega^{D-2} \over (4\pi)^{(D-1)/2}}\sum_{n=0}^\infty\left[
{a_n(\lambda)
\over \Gamma\left({D-1 \over 2}-n\right)}
\omega^{-2n}+
{b_n (\lambda)
\over \Gamma\left({D-2 \over 2}-n\right)}
\omega^{-(2n+1)}\right]~~~.
\end{equation}
In the Euclidean theory one has analogous formulas
\begin{equation}\label{as3}
\mbox{Tr} e^{-t\breve{H}^2(\lambda)}
\simeq {1 \over (4\pi t)^{(D-1)/2}}
\sum_{n=0}^\infty\left[\breve{a}_n(\lambda) t^n
+ \breve{b}_n(\lambda) t^{n+\frac 12}\right]~~~,
\end{equation}
\begin{equation}\label{as4}
\breve{\Phi}(\omega;\lambda)\simeq
{2 \omega^{D-2} \over (4\pi)^{(D-1)/2}}\sum_{n=0}^\infty\left[
{\breve{a}_n(\lambda)
\over \Gamma\left({D-1 \over 2}-n\right)}
\omega^{-2n}+
{\breve{b}_n (\lambda)
\over \Gamma\left({D-2 \over 2}-n\right)}
\omega^{-(2n+1)}\right]~~~.
\end{equation}
The important property of (\ref{as2}) and (\ref{as4}) is that
they are {\it local} functionals of geometrical characteristics
of the background geometry. Thus, if
${\cal M}_E$ and $\cal M$ are related by the "Wick rotation" the
heat kernel coefficients are related by the analytical continuation
\begin{equation}\label{as5}
\breve{a}_n(-iJ,i\lambda)=
a_n(J,\lambda)~~~,~~~
\breve{b}_n(-iJ,i\lambda)=
b_n(J,\lambda)~~~.
\end{equation}
As a consequence, at large $\omega$ the densities coincide, $\breve{\Phi}(\omega,i\lambda|-iJ)=
\Phi(\omega,i\lambda|J)$, and
our second assumption (\ref{ec13}) holds true.
Note that $a_n(\lambda)$,
$\breve{a}_n(\lambda)$ are polynomials
of $\lambda^2$
\begin{equation}\label{as8}
a_n(\lambda)=\sum_{m=0}^{[n/2]}\lambda^{2m}a_{2m,n}~~,~~
\breve{a}_n(\lambda)=\sum_{m=0}^{[n/2]}\lambda^{2m}\breve{a}_{2m,n}~~,
\end{equation}
where $a_{2m,n}$, $\breve{a}_{2m,n}$ are some coefficients (for discussion of (\ref{as8})
see \cite{Fursaev:2000}).
Equations (\ref{as5}) and (\ref{as8}) imply the relation
\begin{equation}\label{as5a}
\breve{a}_{2m,n}(-iJ)=(-1)^m
a_{2m,n}(J)~~~.
\end{equation}
Analogous relations can be found for coefficients $\breve{b}_n$ and
$b_n$.
From (\ref{as4}), (\ref{as8}) one can also get
the following expansion for the density (\ref{ec12})
\begin{equation}\label{as4a}
\breve{\Phi}(\omega)\simeq
{2 \omega^{D-2} \over (4\pi)^{(D-1)/2}}\sum_{n=0}^\infty\left[
{\breve{c}_n
\over \Gamma\left({D-1 \over 2}-n\right)}
\omega^{-2n}+
{\breve{d}_n
\over \Gamma\left({D-2 \over 2}-n\right)}
\omega^{-(2n+1)}\right]~~~,
\end{equation}
where after some algebra one finds
$$
\breve{c}_{n}=\sum_{m=n}^{2n}
{\Gamma\left({D-1 \over 2}-n\right)
\over \Gamma\left({D-1 \over 2}-m\right)}(-1)^{n-m}
\left(\breve{a}_{2(m-n),m}-(m-n+1)\breve{a}_{2(m-n)+2,m+1}\right)
$$
\begin{equation}\label{as4b}
=\sum_{m=n}^{2n}{\Gamma\left(m-{D-1 \over 2}
\right) \over \Gamma\left(n-{D-1 \over 2}\right)}~
\breve{a}_{2(m-n),m}~~~,
\end{equation}
$$
\breve{d}_{n}=\sum_{m=n}^{2n}
{\Gamma\left({D-2 \over 2}-n\right)
\over \Gamma\left({D-2 \over 2}-m\right)} (-1)^{n-m}
\left(\breve{b}_{2(m-n),m}-(m-n+1)\breve{b}_{2(m-n)+2,m+1}\right)
$$
\begin{equation}\label{as4d}
=\sum_{m=n}^{2n}{\Gamma\left(m-{D-2 \over 2}
\right) \over \Gamma\left(n-{D-2 \over 2}\right)}~
\breve{b}_{2(m-n),m}~~~.
\end{equation}
To  derive the last equalities in (\ref{as4b}), (\ref{as4d})
one has to use the properties of the $\Gamma$-function
and also take into account that
$\breve{a}_{2(n+1),2n+1}=\breve{b}_{2(n+1),2n+1}=0$, see
definition (\ref{as8}).
After the "Wick rotation" (\ref{as4a}) coincides precisely with
the asymptotic of the physical density of levels
found in \cite{Fursaev:2000}.

Let us consider now behaviour of the Euclidean action in the
limit of high temperatures when the parameter $\beta$ is small.
We start with (\ref{kk21}), (\ref{kk22})
and replace
$\breve{\Phi}(\omega;\sigma_l)$ by its
asymptotics (\ref{as4}).
(For simplicity we also omit
boundary terms $b_n$.)
If the mass gap is neglected the
integration over $\omega$ can be easily done.
After summation over $\sigma_l\neq 0$,
one finds
\begin{equation}\label{as6}
\zeta(\nu|\beta)\simeq{2\varrho^{-2\nu}\kappa^{D-1-2\nu}
\over (4\pi)^{(D-1)/2} \Gamma(\nu)}
\sum_{n=0}^{\infty}\zeta_R\left(2\nu+2n+1-D\right)
\Gamma\left(\nu+n-{D-1 \over 2}\right)\breve{c}_{n,\nu}\kappa^{-2n}~~~,
\end{equation}
\begin{equation}\label{as7}
\breve{c}_{n,\nu}=\sum_{m=n}^{2n}{\Gamma\left(\nu+m-{D-1 \over 2}
\right) \over \Gamma\left(\nu+n-{D-1 \over 2}\right)}~
\breve{a}_{2(m-n),m}~~~.
\end{equation}
Here $\kappa=2\pi/\beta$, $\zeta_R(z)$ is the Riemann $\zeta$-function
and $\breve{a}_{2m,n}$ are defined by (\ref{as8}).
In principle, one should also add to (\ref{as6}) the term
with $\sigma_l=0$ which we will not discuss here.
By using (\ref{kk21}),(\ref{as6}) and properties
of $\zeta_R(z)$ one gets
corresponding high-temperature expansion for the action
$$
W_E(\beta)\simeq {\kappa^{D-1}
\over (4\pi)^{(D-1)/2} }
\sum_{n=0}^{\infty}\zeta_R\left(2n+1-D\right)
\Gamma\left(n-{D-1 \over 2}\right)\breve{c}_{n,\nu}\kappa^{-2n}
$$
\begin{equation}\label{as9}
=-{1 \over \pi^{D/2}\beta^{D-1}}
\sum_{n=0}^{\infty}\zeta_R\left(D-2n\right)
\Gamma\left({D-2n \over 2}\right)
\breve{c}_{n}
\left({\beta \over 2}\right)^{2n}~~~,
\end{equation}
where, according to (\ref{as4b}),
$\breve{c}_{n}=\breve{c}_{n,\nu=0}$. It should be noted that
one of the terms in this series has a pole at integer $D$. This
corresponds to infrared singularities appearing
if the mass gap is zero.
The nice feature of
(\ref{as9}) is that it is a local functional of the geometry
and the "Wick rotation" from the Euclidean to Lorentzian
geometry here is well defined.
Equation (\ref{as9}) agrees
completely with the high-temperature expansion of the canonical free
energy found in \cite{Fursaev:2000}.

Finally let us remark on
an alternative derivation of (\ref{as9}).
One can start with the following representation of (\ref{kk22})
\begin{equation}\label{as10}
\zeta(\nu|\beta)={\varrho^{-2\nu} \over \Gamma(\nu)}
\int_{0}^{\infty}dt t^{\nu-1} \sum_{\sigma_l}
e^{-\sigma_l^2 t} \mbox{Tr}~e^{-t\breve{H}^2(\sigma_l)}~~~
\end{equation}
and use asymptotic formula (\ref{as3})
for the trace. It enables one to integrate over $t$.
Summation over
$\sigma_l\neq 0$ can be
done if (\ref{as8}) is taken into account. As a result, one gets
(\ref{as6}) for $\zeta(\nu|\beta)$ and (\ref{as9}) for the action.
This way to derive (\ref{as6}), (\ref{as9}) does not depend
on whether the spectrum is continuous or discrete.

\section{Vacuum energy and properties of zeta function}
\setcounter{equation}0

Let us return to the vacuum part $\breve{E}_0$ of the Euclidean action, see
(\ref{ec9}). In static space-times
operators $\breve{H}^2$ and $H^2$ do not depend on
parameter $\lambda$ and coincide.
In this case one gets from
(\ref{ec9}), (\ref{ec3b}) (or (\ref{ec24}))
\begin{equation}\label{v1}
E_0=
\breve{E}_0=-\frac 12\lim_{\nu\rightarrow 0}{d \over d\nu}\zeta_0(\nu)
\end{equation}
\begin{equation}\label{v2}
\zeta_0(\nu)=
{1 \over 2\sqrt{\pi}}
{\Gamma\left(\nu-\frac 12\right) \over \Gamma(\nu)}
\zeta\left(\nu-1/2|\varrho^2 H^2\right)~~.
\end{equation}
$\zeta(z|{\cal O})$ is a generalized $\zeta$-function
of an operator $\cal O$.
Discussion of this definition of the vacuum energy and further
references can be found in \cite{BCVZ:96}.

In stationary space-times, according to
(\ref{ec9}), (\ref{ec3b}) (or (\ref{ec24})),
\begin{equation}\label{v3}
\breve{E}_0=-\frac 12\lim_{\nu\rightarrow 0}{d \over d\nu}\zeta_0(\nu)
\end{equation}
\begin{equation}\label{v4}
\zeta_0(\nu)=
{1 \over \pi\varrho}
\int_0^\infty dx
\zeta\left(\nu|\varrho^2 (\breve{H}^2(x)+x^2)\right)~~.
\end{equation}
Let us show that after the "Wick rotation" $\breve{E}_0$ agrees with the
standard definition
of the vacuum energy. To this aim we represent (\ref{ec3b}) in the form
\begin{equation}\label{v3a}
\zeta_0(\nu)=-{\varrho^{-2\nu} \over 2\pi}\int^{\infty}_{\mu} d\omega \int_{C_+}dz
\breve{\Phi}(\omega;z)(z^2+\omega^2)^{-\nu}e^{i\epsilon z}~~~.
\end{equation}
We have introduced in (\ref{v3a})
a small positive parameter $\epsilon$ to regularize the integral in the limit of
$\nu$ vanishing. By integrating by parts and neglecting terms linear in $\epsilon$
we get
\begin{equation}\label{v3b}
\zeta_0(\nu)=-{\nu \varrho^{-2\nu} \over \pi}\int^{\infty}_{\mu} d\omega \int_{C_+}dz
{2z^2 \over (z^2+\omega^2)^{\nu+1}}
\breve{\Psi}(\omega;z)e^{i\epsilon z}~~~,
\end{equation}
where $\breve{\Psi}(\omega;z)$ is defined by (\ref{ec5}).
The regularization enables us to replace $C_+$ by a closed contour in the upper half of complex
plane and use the Cauchy theorem. According to our first assumption of Section 4, the poles
in the integrand in (\ref{v3b})
are determined only by zeros of the  denominator. So we get
\begin{equation}\label{v3c}
\breve{E}_0
=\frac 12 \int_{\mu}^{\infty} d\omega
\omega \breve{\Phi}(\omega)e^{-\epsilon \omega}~~~.
\end{equation}
Quantity $\breve{\Phi}(\omega)$ is defined in (\ref{ec12}) and corresponds
to the physical spectral density.
After the
"Wick rotation" (\ref{v3c}) turns into the physical vacuum energy with a certain
cutoff $1/\epsilon$ in the range of high frequencies.
Note that our consideration concerns the case when $\breve{B}=1$ (the space-time
is "ultrastationary"). For $\breve{B}\neq 1$   one can use a conformal transformation
to reduce the metric to the ultrastationary form. The vacuum energy in this
case acquires an additional term due to the conformal anomaly, see, e.g.,
\cite{Dowker:89}.

Equations (\ref{v3}), (\ref{v4}) hold for discrete and continuous spectra and
are an important consequence of our analysis.
In principle, they enable one to compute the vacuum
energy of fields in arbitrary stationary space-times
by using the $\zeta$-function method and, thus,  have a number
of applications. In the present paper we restrict the discussion by
some mathematical aspects of the vacuum energy, namely
by analytical properties of the $\zeta$ function (\ref{v4}).
Let us remind that we started with the
$\zeta$ function
$\zeta(\nu|\beta)=\zeta\left(\nu|\varrho^2 L\right)$
of the Euclidean operator $L=-\nabla^2+V$, see (\ref{kk13}), (\ref{kk22}).
According to the general theory, $\zeta(\nu|\beta)$ is a meromorphic
function which has simple poles on the real axis of $\nu$.
In the theory with $D$ dimensions
the part $\zeta^{(p)}(\nu|\beta)$
which includes the poles of $\zeta(\nu|\beta)$
can be written as \cite{BCVZ:96}
\begin{equation}\label{v5}
\zeta^{(p)}(\nu|\beta)={2 \over (4\pi)^{D/2} \Gamma(\nu)}
\sum_{n=0}^\infty {A_n \over 2\nu+2n-D}~~~,
\end{equation}
where $A_n$ are the coefficients related to the asymptotic
of the heat kernel
\begin{equation}\label{v6}
\mbox{Tr}e^{-tL}\simeq{1 \over (4\pi t)^{D/2}}
\sum_{n=0}^\infty A_n t^n~~~.
\end{equation}
In what follows we put $\varrho=1$ for simplicity.
We will also assume that the space-time has no boundaries
and hence $n$ in (\ref{v6}) take only integer values.

\bigskip

The poles of the $\zeta$ function determine
ultraviolet divergences of the Euclidean effective action.
When one separates the action onto the vacuum part $\breve{E}_0$
and the free energy the divergences appear in $\breve{E}_0$.
The free energy has no divergences because of the exponential
factor $e^{-\beta \omega}$ which ensures convergence
of the integral at large $\omega$.
By using these arguments one concludes that
the poles of $\zeta(\nu|\beta)$
and $\beta\zeta_0(\nu)$ coincide.
To find the poles of $\zeta_0(\nu)$ we rewrite (\ref{v4})
as follows
\begin{equation}\label{v7}
\zeta_0(\nu)=
{1 \over \pi \Gamma(\nu)}
\int_0^\infty dx \int_0^\infty dt t^{\nu-1}
\mbox{Tr}~e^{-t(\breve{H}^2(x)+x^2)}~~.
\end{equation}
Our assumption is that the poles of $\zeta_0$ are related
to the behaviour of the integral at small $t$.
We will not present here a rigorous proof of this assumption
but make a check of its consequences for some cases.
By following to \cite{BCVZ:96}
we define the pole part $\zeta^{(p)}_0(\nu)$ of $\zeta_0(\nu)$ by
(\ref{v7}) with integration over $t$ taken in the interval
$(0,1)$. We then replace the trace of the heat kernel of
$\breve{H}^2(x)$ by its asymptotic (\ref{as3})
\begin{equation}\label{v8}
\zeta^{(p)}_0(\nu)=
{1 \over \pi \Gamma(\nu)}
\int_0^\infty dx \int_0^1 dt t^{\nu-1}
e^{-tx^2}{1 \over (4\pi t)^{(D-1)/2}}\sum_{n=0}^\infty \breve{a}_n(x)
t^n~~.
\end{equation}
The integral exists at $\mbox{Re}~\nu>(D-1)/2$.
We can use now (\ref{as8}), integrate over $x$ and then over $t$
to get
\begin{equation}\label{v9}
\zeta^{(p)}_0(\nu)=
{2 \over  (4\pi)^{D/2}\Gamma(\nu)}
\sum_{n=0}^\infty\sum_{m=0}^{[n/2]}
{\Gamma(m+1/2)\breve{a}_{2m,n} \over \sqrt{\pi}(2\nu+2(n-m)-D)}~~.
\end{equation}
The latter equation can be rewritten also as
\begin{equation}\label{v10}
\zeta^{(p)}_0(\nu)=
{2 \over  (4\pi)^{D/2}\Gamma(\nu)}
\sum_{n=0}^\infty {1 \over 2\nu+2n-D}
\sum_{m=n}^{2n}
{\Gamma(m-n+1/2) \over \sqrt{\pi}}
\breve{a}_{2(m-n),m}~~~.
\end{equation}
By comparing poles of (\ref{v5}) and (\ref{v10})  we
conclude that
\begin{equation}\label{v11}
A_n={\beta  \over \sqrt{\pi}}
\sum_{m=n}^{2n} \Gamma(m-n+1/2)
\breve{a}_{2(m-n),m}~~~.
\end{equation}
For $D$ odd this conclusion is true for all $n$ while for $D$ even only
for $0\leq n \leq D/2-1$ because terms with other $n$
do not result in poles. It is clear, however, that (\ref{v11})
should be universal for all $D$ and all $n$ because dimensionality
does not appear in it explicitly.

In some sense (\ref{v11}) can be considered as a
Kaluza-Klein reduction formula for the heat kernel coefficients
of an operator on a stationary $D$ dimensional manifold.
Relation (\ref{v11}) has an interesting and
important consequence.
By comparing it with (\ref{as4b})
for even $D$ and $n=D/2$ one finds that
\begin{equation}\label{v12}
A_{D/2}=\beta \breve{c}_{D/2} ~~~.
\end{equation}
Hence coefficient $\breve{c}_{D/2}$ which appears in the series
(\ref{as4a}), (\ref{as9})
turns out to be related to
the conformal anomaly of the Euclidean theory
\begin{equation}\label{v13}
\zeta(0|\beta)={A_{D/2} \over (4\pi)^{D/2}}
=\beta {\breve{c}_{D/2} \over (4\pi)^{D/2}}
~~~.
\end{equation}
It is not difficult to check the validity of
(\ref{v11}), (\ref{v12}) in some
cases. Of course they true in static space-times and
for $n=0$ in general. Let us consider
(\ref{v11}) for $n\neq 0$
on a $D$-dimensional stationary Euclidean
background\footnote{We remind that our
preceding analysis
concerned space-times of the form
(\ref{i6}) with $\breve{B}=1$.}
\begin{equation}\label{v14}
ds^2 =g_{\mu\nu}dx^\mu dx^\nu
=(d\tau+a_kdx^k)^2+h_{kj}dx^kdx^j~~~.
\end{equation}
Here $\tau$ is periodic with period $\beta$.
We denote this space by ${\cal M}_E$. The $D-1$-dimensional
space $\cal B$
corresponding
to operators $\breve{H}^2(x)$ is determined by the metric
\begin{equation}\label{v15}
dl^2 =h_{kj}dx^kdx^j~~~.
\end{equation}
We denote the Riemann tensors on ${\cal M}_E$ and $\cal B$
as $R_{\mu\nu\lambda\rho}$ and $\bar{R}_{\mu\nu\lambda\rho}$,
respectively. Their relation is discussed in Appendix.

Consider (\ref{v11}) for $n=1$. By using (\ref{v18}) of Appendix
one can find that in any dimension
\begin{equation}\label{v19}
A_1=\beta \int \sqrt{h}d^{D-1}x\left(\frac 16 R[g]-V\right)=
\beta \int \sqrt{h}d^{D-1}x\left(\frac 16 R[h]-{1 \over 24}
F^{\alpha\beta}F_{\alpha\beta}-V\right)
~~~.
\end{equation}
On the other hand, according to (\ref{v11})
\begin{equation}\label{v20}
A_1=\beta \left(\breve{a}_{0,1}+\frac 12 \breve{a}_{2,2}\right)~~~.
\end{equation}
By using heat kernel asymptotics in the external gauge field
one finds
\begin{equation}\label{v21}
\breve{a}_{0,1}=
\int \sqrt{h}d^{D-1}x\left(\frac 16 R[h]-V\right)~~,~~
\breve{a}_{2,2}=-
\int \sqrt{h}d^{D-1}x{1 \over 12}F^{\alpha\beta}F_{\alpha\beta}~~.
\end{equation}
Hence, the right hand sides of (\ref{v19}) and (\ref{v20})
do coincide.

Equation  (\ref{v11}) can be also checked for $n=2$.
In this case it is more non-trivial because $A_2$ is
quadratic in curvatures. The corresponding expression
of the right hand side of (\ref{v11}) contains terms which are quadratic and linear in
curvatures of $\cal B$. It also includes second and forth powers of
the Maxwell tensor  of the fiducial gauge field.
The fomulas become lengthy so we leave all the details for the
Appendix.

\section{Summary}
\setcounter{equation}0

The aim of this paper was to investigate the connection between Euclidean
gravity and statistical mechanics at the one-loop order.
We formulated two assumptions about analytical properties
of the spectral density of the Euclidean single-particle Hamiltonians
$\breve{H}(\lambda)$
which guarantee that the Euclidean one-loop effective action transforms
under the "Wick rotation" to the statistical-mechanical free energy.
We have then showed that our assumptions hold true at least at high
frequencies which determine the high-temperature regime
and ultraviolet properties of the theory.

The important feature of our work is that we posed the question about
Euclidean gravity and thermodynamics in the most general and non-trivial
situation, i.e., when the background gravitational field is stationary and
when the "Wick rotation" implies complexification of the background
metric. To our knowledge, in such a general context, this problem
has not been approached earlier.

As was pointed out, our results can be considered as a consistency check of the Kaluza-Klein
method of \cite{Fursaev:2000} which enables one to deal  with quantum effects
in stationary  gravitational fields. In Euclidean theory
this method has its own advantages and gives new results.
In particular, we have found $\zeta$ function representation, Eqs. (\ref{v3}),
(\ref{v4}), for the average vacuum energy of a scalar field in a stationary
background. Another new result is the Kaluza-Klein reduction formulas
for the heat-kernel coefficients (Eq. (\ref{v11})) and for conformal anomaly
(Eqs. (\ref{v12}), (\ref{v13})).   These results can be used in applications,
especially in the case when one is interested in quantum effects
related to the rotation of the system with respect to the local Lorentz frame.

There are a number of problems for future research.
First of all, we left aside perhaps the most difficult question about analytical
properties of
$\breve{\Phi}(\omega,\lambda|J)$ and $\breve{\Psi}(\omega,\lambda|J)$
at finite frequencies where the theory becomes essentially non-local.
We considered scalar fields, so fields with non-zero spins
require additional analysis. Finally, we dealt with gravitational backgrounds only.
In physical applications, like particles moving in colliders, one also has to allow
external electro-magnetic fields. Inclusion of a background  Abelian gauge field and
real charges in our analysis would be an interesting problem.

\bigskip
\vspace{12pt} {\bf Acknowledgements}:\ \ The work of D.F. is
supported in part by the RFBR grant N 01-02-16791 and NATO
Collaborative Linkage Grant, CLG.976417. A.Z. is grateful to the
Yukawa Institute for Theoretical Physics for hospitality and
financial support. A.Z. is also grateful to the Killam Trust for
its financial support.

\newpage
\appendix
\section{On KK reduction of $A_1$ and $A_2$}
\setcounter{equation}0

To prove (\ref{v11}) for $n=1,2$ we need to use some
geometrical relations between the curvature tensors on
Euclidean $D$--dimensional space ${\cal M}_E$ with metric
\begin{equation}\label{a1}
ds^2 =g_{\mu\nu}dx^\mu dx^\nu
=(d\tau+a_kdx^k)^2+h_{kj}dx^kdx^j~~~,
\end{equation}
where $i,j=1,..D-1$,
and $D-1$-dimensional space $\cal B$ with metric
\begin{equation}\label{a2}
dl^2 =h_{kj}dx^kdx^j~~~.
\end{equation}
The two metrics are related as follows
$g_{\mu\nu}=h_{\mu\nu}+u_\mu u_\nu$.
In the chosen coordinates (\ref{a1})
$u^\mu=(1,0,...0)$, $a_i=-u_i$
and the components $h_{t\mu}=0$.
For space ${\cal M}_E$ vector $u^\mu$ is the Killing vector,
\begin{equation}\label{a2a}
u_{\mu;\nu}+u_{\nu;\mu}=0~~~.
\end{equation}
(Note that (\ref{a2a}) imposes no constraints on the gauge potential
$a_i$.)
It then follows that
\begin{equation}\label{a3}
R_{\lambda\mu\nu\rho}u^\lambda=u_{\mu;\nu\rho}-u_{\mu;\rho\nu}=
-\frac 12 F_{\nu\rho;\mu}~~~,
\end{equation}
where $R_{\mu\nu\lambda\rho}$ is the Riemann tensor on ${\cal M}_E$, and
\begin{equation}\label{a4}
F_{\mu\nu}=2u_{\mu;\nu}~~,~~h^\lambda_\mu F_{\lambda\nu}=F_{\mu\nu}~~,
\end{equation}
is the "Maxwell tensor". From (\ref{a3}) one also gets
another equality
\begin{equation}\label{v22}
R_{\lambda\mu\sigma\rho}u^\lambda u^\sigma=
\frac 14 F^\nu~_\mu F_{\nu\rho}~~~.
\end{equation}

Let us denote curvatures of $\cal B$ with a bar.
It is not difficult to show (see, e.g., \cite{Fursaev:2000})
that the curvatures on ${\cal M}_E$ and $\cal B$ are related
as
\begin{equation}\label{v16}
\bar{R}_{\alpha\beta\gamma\delta}=
h_\alpha^\mu h_\beta^\nu h_\gamma^\lambda h_\delta^\rho
R_{\mu\nu\lambda\rho}+\frac 14 F_{\alpha\gamma}F_{\beta\delta}
-\frac 14 F_{\alpha\delta}F_{\beta\gamma}+\frac 12
F_{\alpha\beta}F_{\gamma\delta}~~~,
\end{equation}
\begin{equation}\label{v17}
\bar{R}_{\alpha\beta}=
h_\alpha^\mu h_\beta^\nu R_{\mu\nu}+\frac 12
F_{\lambda\alpha} F^\lambda~_\beta~~~,
\end{equation}
\begin{equation}\label{v18}
\bar{R}=R+
\frac 14 F^{\alpha\beta}F_{\alpha\beta}~~~.
\end{equation}
$h_\alpha^\mu$ is the projector on the plane orthogonal to $u^\mu$.
Relation (\ref{v17}) follows from
(\ref{v16}), (\ref{v22}).

Equations (\ref{v16})--(\ref{v18}) can be used
to find representation for invariants quadratic in curvature tensors.
By using (\ref{v16}), (\ref{a3}), (\ref{v22})
and properties of the Riemann tensor
one finds after some algebra
$$
R^{\lambda\mu\nu\rho}R_{\lambda\mu\nu\rho}=
\bar{R}^{\lambda\mu\nu\rho}\bar{R}_{\lambda\mu\nu\rho}
-\frac 32\bar{R}^{\lambda\nu\mu\rho}F_{\lambda\nu}F_{\mu\rho}
$$
\begin{equation}\label{a5}
+\frac 38 (F^{\mu\nu}F_{\mu\nu})^2
+F_{\nu\rho;\mu}F^{\nu\rho;\mu}+
\frac 18 F^{\lambda\nu}F^{\mu\rho}F_{\lambda\mu}F_{\nu\rho}~~~.
\end{equation}
This relation can be further transformed if we note that
$$
F_{\nu\rho;\mu}F^{\nu\rho;\mu}=-2
F_{\nu\rho;\mu}F^{\rho\mu;\nu}=
2F_{\mu\rho}~^{;\rho}F^{\mu\lambda}~_{;\lambda}
$$
\begin{equation}\label{a6}
+R^{\lambda\nu\mu\rho}F_{\lambda\nu}F_{\mu\rho}-2R_{\mu\nu}
F^{\rho\mu}F_\rho~^\nu-2(F_{\rho}~^\nu F^{\rho\mu})_{;\nu\mu}
+(F_{\nu\rho}F^{\nu\rho})^{;\mu}~_{\mu}
~~~,
\end{equation}
$$
R^{\lambda\nu\mu\rho}F_{\lambda\nu}F_{\mu\rho}-2R_{\mu\nu}
F^{\rho\mu}F_\rho~^\nu=
$$
\begin{equation}\label{a7}
\bar{R}^{\lambda\nu\mu\rho}F_{\lambda\nu}F_{\mu\rho}-2\bar{R}_{\mu\nu}
F^{\rho\mu}F_\rho~^\nu
-\frac 12 (F_{\mu\nu}F^{\mu\nu})^2
+\frac 12 F^{\lambda\nu}F^{\mu\rho}F_{\lambda\mu}F_{\nu\rho}
~~~,
\end{equation}
and, hence, that
$$
F_{\nu\rho;\mu}F^{\nu\rho;\mu}=
2F_{\mu\rho}~^{;\rho}F^{\mu\lambda}~_{;\lambda}+
\bar{R}^{\lambda\nu\mu\rho}F_{\lambda\nu}F_{\mu\rho}
$$
\begin{equation}\label{a8}
-2\bar{R}_{\mu\nu}F^{\rho\mu}F_\rho~^\nu
-\frac 12 (F_{\mu\nu}F^{\mu\nu})^2
+\frac 12 F^{\lambda\nu}F^{\mu\rho}F_{\lambda\mu}F_{\nu\rho}
-2(F_{\rho}~^\nu F^{\rho\mu})_{;\nu\mu}
+(F_{\nu\rho}F^{\nu\rho})^{;\mu}~_{\mu}
~~~.
\end{equation}
Thus, the final result for the square of the Riemann tensor (\ref{a5})
is
$$
R^{\lambda\mu\nu\rho}R_{\lambda\mu\nu\rho}=
\bar{R}^{\lambda\mu\nu\rho}\bar{R}_{\lambda\mu\nu\rho}
-\frac 12\bar{R}^{\lambda\nu\mu\rho}F_{\lambda\nu}F_{\mu\rho}
-2\bar{R}_{\mu\nu}F^{\rho\mu}F_{\rho}~^\nu
-\frac 18 (F^{\mu\nu}F_{\mu\nu})^2
$$
\begin{equation}\label{a9}
+\frac 58 F^{\lambda\nu}F^{\mu\rho}F_{\lambda\mu}F_{\nu\rho}
+2F_{\mu\rho}~^{;\rho}F^{\mu\lambda}~_{;\lambda}
-2(F_\rho~^\nu F^{\rho\mu})_{;\nu\mu}
+(F_{\nu\rho}F^{\nu\rho})^{;\mu}~_{\mu}
~~.
\end{equation}
In a similar way
by using (\ref{v17}), (\ref{a3}) and (\ref{v22}) one gets
the following result for the square of the Ricci tensor

$$
R^{\mu\nu}R_{\mu\nu}=\bar{R}^{\mu\nu}\bar{R}_{\mu\nu}
$$
\begin{equation}\label{a10}
-\bar{R}_{\mu\nu}F^{\rho\mu}F_{\rho}~^\nu+
\frac 14 F^{\lambda\nu}F^{\mu\rho}F_{\lambda\mu}F_{\nu\rho}
-{1 \over 16} (F^{\mu\nu}F_{\mu\nu})^2
+\frac 12 F_{\mu\rho}~^{;\rho}F^{\mu\lambda}~_{;\lambda}~~.
\end{equation}
Finally, from (\ref{v18}) one gets for the square of the
scalar curvature
\begin{equation}\label{a11}
R^2=\bar{R}^2
-{1 \over 2} (F^{\mu\nu}F_{\mu\nu})\bar{R}
+{1 \over 16} (F^{\mu\nu}F_{\mu\nu})^2~~~.
\end{equation}

The coefficient $A_2$ on a closed manifold can be written  by using
(\ref{a9})--(\ref{a11}) as
$$
A_2=
\beta \int \sqrt{h}d^{D-1}x\left(
{1 \over 72} R^2-{1 \over 180} R_{\mu\nu}R^{\mu\nu}
+{1 \over 180} R_{\mu\nu\lambda\rho}R^{\mu\nu\lambda\rho} -\frac 16 RV +\frac 12 V^2
\right)=
$$
$$
\beta \int \sqrt{h}d^{D-1}x\left(
{1 \over 72} \bar{R}^2-{1 \over 180}
\bar{R}_{\mu\nu}\bar{R}^{\mu\nu}
+{1 \over 180} \bar{R}_{\mu\nu\lambda\rho}\bar{R}^{\mu\nu\lambda\rho}
\right.
$$
$$
+{1 \over 120} F_{\mu\rho}~^{||\rho}F^{\mu\lambda}~_{||\lambda}
-{1 \over 144} (F^{\mu\nu}F_{\mu\nu})\bar{R}
-{1 \over 360}\bar{R}^{\lambda\nu\mu\rho}F_{\lambda\nu}F_{\mu\rho}
-{1 \over 180}\bar{R}_{\mu\nu}F^{\rho\mu}F_{\rho}~^\nu
$$
\begin{equation}\label{a12}
\left.+{1 \over 384} (F^{\mu\nu}F_{\mu\nu})^2
+{1 \over 480}
F^{\lambda\nu}F^{\mu\rho}F_{\lambda\mu}F_{\nu\rho}\right)~~~.
\end{equation}
Here the symbol $||$ corresponds to covariant
differentiation in the metric $h_{\mu\nu}$ and to get (\ref{a12})
we took into account the identity
$$
F_{\mu\rho}~^{||\rho}=F_{\mu\rho}~^{;\rho}+\frac 12 u_\mu
F^{\lambda\nu}F_{\lambda\nu}~~~.
$$
Relation (\ref{v11}) for $n=2$  is
\begin{equation}\label{a13}
A_2=\beta \left(\breve{a}_{0,2}+\frac 12 \breve{a}_{2,3}
+\frac 34 \breve{a}_{4,4}\right)~~~.
\end{equation}
The coefficient $\breve{a}_{0,2}$,
$\breve{a}_{2,3}$, $\breve{a}_{4,4}$
can be found by using
results of \cite{Avramidi}, \cite{BGV}.
\begin{equation}\label{a15}
\breve{a}_{0,2}=
\int \sqrt{h}d^{D-1}x\left(
{1 \over 72} \bar{R}^2-{1 \over 180}
\bar{R}_{\mu\nu}\bar{R}^{\mu\nu}
+{1 \over 180} \bar{R}_{\mu\nu\lambda\rho}\bar{R}^{\mu\nu\lambda\rho}-\frac 16 \bar{R}V
+\frac 12 V^2
\right)~~,
\end{equation}
$$
\breve{a}_{2,3}=\int \sqrt{h} d^{D-1}x\left({1 \over 12} F^{\mu\nu}F_{\mu\nu} V    \right.
$$
\begin{equation}\label{a16}
+ \left.
{1 \over 60} F_{\mu\rho}~^{||\rho}F^{\mu\lambda}~_{||\lambda}
-{1 \over 72} (F^{\mu\nu}F_{\mu\nu})\bar{R}
-{1 \over 180}\bar{R}^{\lambda\nu\mu\rho}F_{\lambda\nu}F_{\mu\rho}
-{1 \over 90}\bar{R}_{\mu\nu}F^{\rho\mu}F_{\rho}~^\nu \right)~~,
\end{equation}
\begin{equation}\label{a17}
\breve{a}_{4,4}=\int \sqrt{h} d^{D-1}x \left(
{1 \over 288} (F^{\mu\nu}F_{\mu\nu})^2
+{1 \over 360}
F^{\lambda\nu}F^{\mu\rho}F_{\lambda\mu}F_{\nu\rho}\right)~~~.
\end{equation}
By using these results one can check that (\ref{a13}) holds true.

\end{document}